\documentstyle[preprint,aps]{revtex}
\begin{document}

\newcommand\beq{\begin{equation}}
\newcommand\eeq{\end{equation}}
\newcommand\bea{\begin{eqnarray}}
\newcommand\eea{\end{eqnarray}}
\tightenlines
\draft
\title{Solitons in a one-dimensional interacting Bose-Einstein system}
\author{R. K. Bhaduri$^1$, Sankalpa Ghosh$^2$, M. V. N. Murthy$^2$ and 
Diptiman Sen$^3$} 
\address{1. Department of Physics and Astronomy, McMaster University,
Hamilton, Ontario, Canada L8S 4M1}
\address{2. The Institute of Mathematical Sciences, Chennai 600 113, India}
\address{3. Center for Theoretical Studies, Indian Institute of Science, 
Bangalore 560 012, India}
\maketitle

\begin{abstract}

A modified Gross-Pitaevskii approximation was introduced recently for
bosons in dimension $d\le2$ by Kolomeisky {\it et al.} (Phys. Rev. Lett.
{\bf 85} 1146 (2000)). We use the density functional approach with 
sixth-degree interaction energy term in the Bose field to reproduce the
stationary-frame results of Kolomeisky {\it et al.} for a one-dimensional
Bose-Einstein system with a repulsive interaction. We also find a soliton
solution for an attractive interaction, which may be boosted to a finite
velocity by a Galilean transformation. The stability of such a soliton is
discussed analytically. We provide a general treatment of stationary
solutions in one dimension which includes the above solutions as special
cases. This treatment leads to a variety of stationary wave solutions for
both attractive and repulsive interactions. 

\end{abstract}
\vskip .5 true cm

\maketitle
\narrowtext

\section{Introduction}

There has been a number of theoretical studies on Bose-Einstein
condensation (BEC) in a trap that is effectively lower dimensional
\cite{bagnato,ruprecht,ketterle,perez,muryshev,carr,petrov,bhaduri}. A
three-dimensional (axially symmetric) anisotropic harmonic trap may be
regarded as (quasi) two- or one-dimensional when $\omega_{\bot}\ll
\omega_x$, or $\omega_{\bot}\gg \omega_x$ respectively, at temperatures
much smaller than the larger of the oscillator energy gaps. Recently,
Burger {\it et al.} \cite{burger} have experimentally created dark
solitons in such a quasi one-dimensional system of dilute $Rb^{87}$ atoms.
In this experiment, $\omega_{\bot}$ was taken to be about $30$ times
larger than $\omega_x$. After the creation of the soliton by
phase-imprinting on the cigar-shaped condensate, the trap is switched off,
and the soliton is observed to propagate along the $x$-direction. The
$s$-wave inter-atomic interaction between $Rb^{87}$ atoms is repulsive,
and stable dark solitons are expected to exist in one dimension, as seen
in this experiment. We note, however, that for $Rb^{85}$ and $Li^{7}$, the
interatomic interaction is attractive, and bright solitons may form. Such
solitons have been studied for the quasi one-dimensional case by several
groups \cite{perez,carr}, using the Gross-Pitaevskii (GP) mean-field
theory \cite{gp}. The standard GP theory replaces the bosonic field
operator by a classical order parameter field $\Phi({\bf r},t)$ in three
dimensions, and the potential energy is given by a fourth-degree
$|\Phi|^4$ term in the energy-density functional. Perez-Garcia {\it et
al.} \cite{perez} have shown how this may be reduced, for extreme
anisotropy ($\omega_{\bot} \gg \omega_x$), to the quasi one-dimensional
geometry with a fourth-degree potential energy term. It has been also
shown that for the one-dimensional case with this attractive fourth-degree
interaction potential, the solitons are dynamically stable ( for the
repulsive case, this has been known for a long time \cite{zakharov}). One
has also considered the limiting case for $\omega_x=0$, when there is no
confinement in the $x$-direction, and studied the free propagation of the
soliton. 

A different approach has been taken recently by Kolomeisky {\it et al.}
\cite{kolo}. Instead of starting with the usual three-dimensional GP
energy functional and reducing it to the quasi one-dimensional case as
described above, they have derived the GP functional for truly lower
dimensions, $d\le 2$, using the renormalization group approach
\cite{kolo2}. In particular, they proposed, for a repulsive interatomic
interaction, a sixth-degree potential energy term replacing the usual
fourth-degree term in the order parameter. With this modification they
obtained a (dark) soliton solution in the repulsive case. 

In this paper, we first discuss the solutions of the modified GP formalism
in one dimension in the stationary frame. We show that 
the soliton solution of Kolomeisky {\it et al.} \cite{kolo} may be
reproduced in the stationary frame with a repulsive interaction. Since the
number of particles (see the definition in the next section) is generally
not finite unless regulated, the phase cannot be obtained by a Galilean
boost even though the system has Galilean invariance to begin with. 

Next we focus on an attractive sixth-degree interaction term in the energy
functional for bosons. Unlike the repulsive sixth-degree interaction, the
attractive interaction may not be obtainable as a modified GP functional
in one dimension. However, the motivations for taking such a
phenomenological attractive potential energy functional are two-fold. In
three dimensions, using the standard fourth-degree attractive interaction
energy term in the GP functional, it was predicted that the soliton-like
solution would be stable only up to a limited number of bosons in the
condensate \cite{ruprecht,dalfovo,kagan,dodd}, and this was verified
experimentally \cite{bradley}. Of course, in one dimension with an
attractive fourth-degree term, the soliton is variationally stable as we
will show in Sec. III D below. We want to examine the analytical solutions
for the attractive sixth-degree interaction in the one-dimensional case. 
Analytical solutions for a more general case which includes the
sixth-degree interaction have been derived earlier, in a different
context, by Pelinovsky {\it et al.} \cite{pelinovsky}. They have also
discussed numerically the stability of these solutions. Here we use the
density functional approach to derive these analytical solutions in the
stationary frame. We show that there exists a limitation on the particle
number for stability. Since the particle number is finite in the
attractive case, we may use the Galilean invariance to obtain the general
solution in an arbitrary frame. 

It should also be mentioned that the fourth- and sixth-degree interaction
(in the field variable) in the energy functional has been used extensively
in various applications of physics \cite{barashenkov}, with a recent one
\cite{das} examining the stability of the $Li^7$ condensate with an
attractive two-body and a repulsive three-body contact interaction in
three dimensions. More recently, the stability of an attractive BEC in the
presence of a sixth-degree interaction with confinement has been analyzed
by Zhang\cite{zhang}. The purely sixth-degree energy functional in one
dimension, without the confining potential considered by Kolomeisky {\it
et al.} \cite{kolo}, is special because it is scale-invariant, and it has
a peculiar condition on the particle number for the soliton in the
attractive case. We further show that there are soliton as well as
stationary wave solutions if one uses the phenomenological model for an
attractive interaction. 

In Sec. II, we discuss the equations of motion in the stationary frame and
their symmetries. In Sec. III, we obtain the solutions for repulsive and
attractive interactions, and discuss the stability of the soliton solution
in the attractive case. Sec. IV provides a general discussion of the
stationary solutions which includes all the above solutions as special cases. 

\section{Equations of Motion}

Our starting point is Eqs. (3) and (4) of Kolomeisky {\it et al.}
\cite{kolo} without a trap potential. The modified GP functional for a 
one-dimensional Bose-Einstein system is given by 
\beq 
E ~=~ \frac{\hbar^2}{2m} \int dx ~[~ \frac{d\Phi}{dx}\frac{d\Phi^*}{dx} ~
+~ \frac{g\pi^2}{3} ~(\Phi \Phi^*)^3 ~] ~,
\label{GPE} 
\eeq 
The first term is the kinetic energy density and the second term arises from 
interactions; here the dimensionless coupling $g$ may be positive or negative. 
Note that the interaction term in (\ref{GPE}) is cubic in the density $\rho = 
\Phi \Phi^{\star}$, and not quadratic as in the standard theory. A rigorous 
derivation of this term for the repulsive case may be found in 
Ref. \cite{kolo2}. It suffices to state here that in one dimension, where 
particles are not allowed to cross, such a term naturally arises when the 
interaction is short-range. This has been demonstrated for the
hard-core interaction by Girardeau \cite{gira} in the thermodynamic limit, 
and by Lieb and Liniger \cite{lieb} for a repulsive delta function potential. 
Even with a two-body interaction which varies 
inversely as the distance between particles, the leading term in the 
density functional has a cubic dependence to the density \cite{sen2}.

A few remarks may be made here in connection with the comment made by 
Bhaduri and Sen \cite{sen1} recently. The Lieb and Liniger analysis 
\cite{lieb} yields the result that in the limit of very low density of 
very strong repulsive interaction, the system becomes fermionic with
$g=1$ in the functional given in Eq. (\ref{GPE}). In the ground state, 
the density is constant and the kinetic term vanishes. 
Hence one obtains the free fermion result for the energy per unit length,
\beq
E = \frac{\hbar^2}{2m} \frac{\pi^2}{3} \rho^3 ~.
\label{tf}
\eeq
Obviously, in the free fermion limit no soliton solutions occur. 
In what follows, we keep both the kinetic term and the interaction term 
between the bosons with a general coupling $g$, and therefore 
are not in the free fermion limit. 

The equation of motion may be obtained from the energy functional in Eq. 
(\ref{GPE}), and is given by
\beq
{\hbar^{2} \over 2m} ~[~ -~{\partial^{2} \over {\partial x^{2}}} \;
+~ g \pi^2 \; |\Phi|^4 ~]~ \Phi(x,t) \; = \; i\hbar ~{\partial\Phi(x,t)
\over{\partial t}} \;\;.
\label{scheq}
\eeq
Henceforth we shall denote the partial derivatives with respect to $x$
by a prime and with respect to $t$ by a dot.
It is useful to multiply Eq. (\ref{scheq}) by $\Phi^{*}(x,t)$,
\beq
-{\hbar^{2} \over 2m}\Phi^{*}\Phi^{\prime \prime} \; + \; \Phi^{*}W(x)\Phi \;
= \; i\hbar\Phi^{*}\dot\Phi \;\;,
\label{hden1}
\eeq
and take the complex conjugate,
\beq
-{\hbar^{2} \over 2m}\Phi \Phi^{*\prime \prime} \; + \; \Phi W(x)\Phi^{*}
\; = \; -i\hbar \Phi \dot \Phi^{*} \;\;.
\label{hden2}
\eeq
where $W(x) = (\hbar^2 /2m) g \pi^2 |\Phi|^4$. 

We now set, quite generally,
\beq
\Phi(x,t) \; = \; {\sqrt{\rho(x,t)}}\;\;{\rm e}^{i\theta(x,t)} \;\;,
\label{defpsi}
\eeq
where $\rho$ and $\theta$ are real. It is useful to define the total particle 
number
\beq
N ~=~ \int ~dx ~\rho ~,
\eeq
and the momentum functional 
\beq
P ~=~ \frac{-i\hbar}{2} ~\int ~dx ~[~ \Phi^* \Phi^\prime ~-~ 
\Phi^{* \prime} \Phi ~]~ =~ \hbar ~\int ~dx ~\rho \theta^\prime ~. 
\label{mom}
\eeq

We then obtain the equations of motion as follows. The difference between 
Eqs. (\ref{hden1}) and (\ref{hden2}) reduces to the continuity equation
for the number density $\rho$,
\beq
\dot\rho \; + \; {\hbar\over m}~ (\rho\theta^\prime )^\prime \; = \; 0\;\;.
\label{conteq}
\eeq
Thus it is legitimate to regard the velocity of the ``fluid'' to be
\beq
v(x,t) \; = \; {\hbar\over m} ~\theta^\prime (x,t) \;\;.
\label{velocity}
\eeq
The energy density equation is obtained by adding Eqs. (\ref{hden1})
and (\ref{hden2}) and dividing by 2,
\beq
\tau(x,t) \; + \; W(x)\rho \; = \; - \hbar\rho\dot\theta \;\;,
\label{enden}
\eeq
where the kinetic energy density is
\bea
\tau(x,t) \; & = & \; -{\hbar^{2} \over 4m}(\Phi^{*}\Phi^{\prime \prime} 
+\Phi^{* \prime \prime} \Phi) \nonumber \\
& = & \; {\hbar^{2} \over 2m} \left({1\over4}{(\rho^\prime )^{2} \over \rho}
-{1\over2}\rho^{\prime \prime} +\rho(\theta^\prime )^{2}\right) \;\;.
\label{kinetic}
\eea

Note that the Eqs. (\ref{conteq}) and (\ref{enden}) are invariant under 
scaling and Galilean boosts \cite{sen2}. Under scaling by a factor $\alpha$,
\bea
\rho (x,t) ~&\rightarrow &~\alpha \rho (\alpha x, \alpha^2 t) ~, \nonumber \\
\theta (x,t) ~&\rightarrow &~\theta (\alpha x, \alpha^2 t) ~.
\label{scale}
\eea
Then the particle number $N$ remains invariant, while the energy and momentum 
transform as 
\beq
E ~\rightarrow ~\alpha^2 E \quad {\rm and} \quad P ~\rightarrow ~ \alpha 
P ~. 
\eeq
Under a Galilean boost by velocity $v$,
\bea
\rho (x,t) ~&\rightarrow &~ \rho (x -vt, t) ~, \nonumber \\
\theta (x,t) ~&\rightarrow &~\theta (x -vt, t) ~+~ \frac{mv}{\hbar} ~(x ~-~ 
\frac{1}{2} vt) ~.
\label{boost}
\eea
Then the particle number $N$ remains invariant while the energy and momentum 
transform as
\bea
E ~&\rightarrow &~E ~+~ Pv ~+~ \frac{1}{2} mN v^2 ~, \nonumber \\
P ~&\rightarrow &~P ~+~ mN v ~. 
\label{enmom}
\eea
We should mention a subtlety here. If we are interested in solutions involving 
a single localized object (such as the solitons discussed in the next section),
then it is reasonable to demand that its energy and momentum should only 
change by a finite amount under a boost. From Eq. (\ref{enmom}), we see that 
this is true only if the particle number $N$ is finite. Therefore we will 
perform boosts only when $N$ is finite.

\section{Stationary Solutions: Some Analytical Results}

We will now find some stationary solutions of Eqs. (\ref{conteq}) and 
(\ref{enden}). For such solutions we make the following ansatz, 
\beq
\Phi(x,t) ~=~ \sqrt{\rho(x)} ~\exp ~(-i\mu t/\hbar) ~,
\eeq
namely, we set $\theta(x,t) = -\mu t /\hbar$. Note that this ansatz 
is consistent with the continuity equation since both $\dot \rho$ and 
$\theta^\prime$ are zero. The equation for $\rho(x)$ is then given by
\beq
\mu \rho ~=~ {\hbar^{2} \over 2m} \left({1\over4}{(\rho^\prime )^{2} 
\over \rho} -{1\over2}\rho^{\prime \prime}+ g \pi^2 \rho^3\right) \;\;. 
\label{GPDENSE}
\eeq
Then the following solutions exist.

\subsection{Thomas-Fermi limit}

Assuming the bulk density to be a constant, we set
$\rho (x) = \rho_0 $. This is an allowed solution with 
\beq 
\rho_0 ~=~ \sqrt{2m\mu /g\pi^2 \hbar^2},
\label{tfd}
\eeq
provided $g$ is positive. This is the Thomas-Fermi result \cite{kolo}. 
Computing the energy, we reproduce the result given in Eq. (\ref{tf}) when 
$g=1$. 

\subsection {Repulsive interaction}

We assume $g$ is positive and look for solutions when the density is not 
a constant. The Kolomeisky {\it et al.} solution given in their Eq. (12) 
may be written in the stationary frame with their velocity parameter 
$\beta =0$ \cite{kolo}. The solution is given by 
\beq
\rho(x) ~=~ \frac{k}{\pi {\sqrt g}} ~\frac{\cosh (2kx) - 1}{\cosh (2kx) + 2} 
\label{KOMOL}
\eeq
with
\beq
\mu ~=~ \frac{\hbar^2 k^2}{2m} ~.
\eeq
Note that the density tends to $\rho_0 =1$ at $x \rightarrow \pm \infty$;
hence it is better to define the particle number of the soliton to be
\beq
N_{soliton} ~=~ \int ~dx ~[ \rho_0 - \rho (x)] ~=~ \frac{3}{2\pi {\sqrt g}} ~
\int_{-\infty}^{\infty} ~\frac{dt}{\cosh t + 2} ~=~\frac{\sqrt{3}}{\pi 
{\sqrt g}} ~\ln \left[ \frac{\sqrt{3}+1}{\sqrt{3}-1} \right] ~. 
\eeq
In non-linear optics this number corresponds to the total intensity of 
light with a constant background subtracted out. Since 
$N_{soliton}$ is related to $g$, there exists exactly one solution for 
the system for a given value of $g$ (up to a scaling of $k$).

The above solution is called a dark soliton in optics \cite{physrep}
because the density goes to zero at $x=0$, and the soliton physically
corresponds to a dip in the density distribution. Note that the above
solution is valid when $g$ is positive, i.e., for repulsive interactions. 

One can get the energy, $E$ of the soliton by integrating the energy density
\cite{rajaraman} given in Eq. (\ref{GPE}). Using the soliton 
profile given in Eq. (\ref{KOMOL}) and using Eq. (\ref{GPDENSE})
it is easy to see that the energy of the soliton is infinite. However if 
the energy of the soliton is defined as the difference
\beq
E_{soliton} = E -E_0 ~,
\label{SOLITM1}
\eeq
where $E_0$ is the energy of the solution with $\rho(x)=\rho_0=k/\pi {\sqrt 
g}$, then it turns out that the energy of the soliton is zero. It is 
interesting that even in the presence of a repulsive interaction, it 
does not cost any energy to create a soliton. 

Until now we have checked the results given in 
Ref. \cite{kolo} in the stationary frame. Next we consider some new 
solutions in the presence of an attractive interaction. 

\subsection{Attractive interaction}

The equation of motion for $\Phi(x,t)$ in 
the stationary frame for $\rho$ may also be solved by the following ansatz 
\beq
\rho ~=~ \frac{2kN}{\pi} ~{\rm sech} ~(2kx) \quad {\rm and} \quad \theta ~
=~ - ~ \frac{\mu t}{\hbar} ~.
\label{solit2}
\eeq
Note that the density is normalized so that 
\beq
\int_{-\infty}^{\infty} ~dx ~\rho(x) ~=~ N ~.
\eeq
The ansatz in (\ref{solit2}) satisfies the equation of motion provided that
\beq
N^2 ~=~ -\frac{3}{4g} \quad {\rm and} \quad \mu ~=~ -\frac{\hbar^2 k^2}{2m} ~.
\eeq
Since $N$ is real we are forced to choose $g$ to be negative. We may 
therefore write
\beq
N^2 ~=~ \frac{3}{4|g|} ~.
\label{solnum}
\eeq
Following Ref. \cite{physrep} we call this the bright soliton. 
We note in passing that a spatial soliton of the secant hyperbolic type is 
well-known in nonlinear optics, and has been experimentally realized 
long back \cite{chiao}, although the differential equation obeyed by the
solitonic electrical field in optics is very different from ours. 

Since $N$ and $g$ are related to each other by Eq. (\ref{solnum}), there 
exists exactly one solution (up to scaling) for the system given a value for
either $g$ or $N$. The soliton solution in this case is a lump, and it has 
its maximum at $x=0$. Note the peculiar fact that the larger the magnitude 
of $g$, the smaller is the particle number $N$ of the soliton. The reason for
this is explained in the next subsection where we will show that a 
configuration whose particle number is too large is unstable to collapse.

Interestingly, we find that the energy of the soliton in this case is zero.
Since $N$ is finite, we can use Eq. (\ref{boost}), to boost the 
solution given in Eq. (\ref{solit2}). Thus
\bea
\rho ~&=&~ \frac{2kN}{\pi} ~{\rm sech} ~(2k (x-vt)) ~, \nonumber \\
\theta ~&=&~ - ~ \frac{\mu t}{\hbar} ~+~ \frac{mv}{\hbar} (x - \frac{1}{2} 
vt) ~.
\eea
We then find that 
\beq E ~=~ \frac{1}{2} mN v^2 \quad {\rm and} \quad P ~=~ mNv ~.
\eeq
Hence it is natural to define the mass of this soliton to be $M =mN$.

We further note that there also exist stationary wave solutions 
apart from the above. Starting from the dark
soliton solution given in Eq. (\ref{KOMOL}), we can transform $k \rightarrow 
i k$ and $\rho \rightarrow - \rho$ to obtain the following stationary wave 
solution for {\it negative} values of $g$:
\beq
\rho(x) ~=~ \frac{k}{\pi {\sqrt {-g}}} ~\frac{1 - \cos (2kx)}{2 + \cos 
(2kx)} ~, 
\label{wave}
\eeq
with 
\beq
\mu ~=~ - \frac{\hbar^2 k^2}{2m} ~.
\eeq

Finally, note that we can apply the scaling transformation given in Eq. 
(\ref{scale}) for all the solutions given above; that would change the 
normalization of the density in all cases. But we can apply the boost
transformation only for the bright soliton since that is the only case in 
which the total particle number is finite.

\subsection{Stability of the soliton solutions}

We will now study the stability of the solitons, particularly the bright 
soliton given in Eq. (\ref{solit2}), using both variational and small 
fluctuation analysis. Before doing that, let us make some preliminary comments.

The soliton solutions obtained above for both attractive and repulsive
cases are scale invariant. Hence the type of instability which is
associated with the soliton solutions for certain types of nonlinear
Schr\"odinger equations (see Barashenkov {\it et al.} \cite{bara})
are not expected here. Another form of instability
which can occur is the breaking of the soliton into some other fundamental
constituents \cite{raj}. To check the possibility of that, let us note that 
our energy functional has a global $U(1)$ symmetry, and hence a conserved 
charge associated with such a soliton solution. For the attractive case, 
the charge is given by the expression 
\beq 
Q ~= ~i\hbar
~\int dx ~(\Phi^{\dag}\dot \Phi - \dot\Phi^{\dag}\Phi) ~=~ 2 \mu N ~=~
-~\frac{\hbar^2 k^2}{2m} ~\sqrt{\frac{3}{|g|}} ~. 
\eeq 
This is of course proportional to the integral of the density $\rho(x)$ in
our case which is the same in any frame. In that case the soliton solution
is classically stable against the aforesaid complete
dissociation into more fundamental constituents (mfc), because its rest
energy which is $0$ here must be lower than $m_0 |Q|$ if we assume that
the mass $m_0$ of the hypothetical mfc must be positive. 

We will now show that in the attractive case $g<0$, the bright soliton can 
collapse to a point if the particle number of a configuration is bigger than 
a critical number $N_c$ which is a function of the parameter $g$. We 
can show this as follows. Consider a trial wave function of the form
\beq
\Phi (x) ~=~ \left( \frac{N\alpha}{\pi \cosh (\alpha x)} \right)^{1/2} ~,
\label{var}
\eeq
where $\alpha$ is a free parameter (whose inverse determines the 
width of the configuration), and we set $\theta =0$.
The particle number for this configuration is given by $N$. We then find 
that the energy functional in Eq. (1) is given by
\beq
E ~=~ \frac{\alpha^2 \hbar^2 N}{16m} ~[~ 1 ~-~ \frac{4}{3} ~|g| N^2 ~]~.
\eeq
for this configuration. We observe that this expression does not have a 
minimum at any finite and nonzero value of $\alpha$. If 
\beq
N^2 ~>~ N_c^2 ~\equiv ~ \frac{3}{4|g|} ~,
\label{coll}
\eeq
then the energy of the configuration given in Eq. (\ref{var})
can be made arbitrarily large and negative by letting 
$\alpha \rightarrow \infty$; this corresponds to the density of the 
configuration collapsing to a $\delta$-function at the origin. We 
thus see that there is a critical value of $N_c$ which is proportional 
to $1/{\sqrt {|g|}}$ beyond which there is no lower bound to the 
energy. We observe that the particle number of the bright soliton in Eq. 
(\ref{solnum}) is exactly equal to this critical value. Since this argument
is only based on a trial wave function, it does not prove anything about the 
stability of the bright soliton against a collapse.

It is instructive to contrast the situation here with a more general
form of the GP functional where the interaction term in
Eq. (\ref{GPE}) is taken to be of the form $g \rho^n$ instead of $g \rho^3$,
where we assume $n>1$. If the exponent of $\rho$ is taken to be $n<3$ 
(instead of $n=3$ as we have considered so far), and $g$ is negative, we find 
from the above argument that there is no instability to a collapse, no matter 
what the values of $g$ and $N$ are. Namely, for any value of $g <0$ and $N$, 
the energy of the trial configuration given in (\ref{var}) has a minimum at 
some finite value of the inverse width $\alpha$. On the other hand, if we 
take the exponent $n>3$ and $g<0$, the above argument indicates an instability
to collapse for any value of $N$, no matter how small. Thus, an attractive
interaction of the form $\rho^3$ is rather special; this is the only value of 
the exponent $n$ for which there is a finite critical number $N_c$, such that
the trial configuration is
stable against a collapse if $N<N_c$ and is unstable if $N>N_c$. This special 
property of $n=3$ can be traced to the fact that the coefficient of the 
interaction $g$ is dimensionless only for this value of $n$.

Let us now return to the bright soliton given in Eq. (\ref{solit2}), and
directly study its stability by studying small fluctuations around it. Our
main conclusion will be that there is at least one unstable mode. To do the 
analysis, let us set $k=1/2$ for simplicity in (\ref{solit2}). Then the 
soliton takes the form
\bea
\rho_c &=& \frac{N}{\pi \cosh x} ~, \nonumber \\
\theta_c &=& \frac{\hbar t}{8m} ~,
\eea
with $N^2 g = -3/4$. We now assume small fluctuations of the form
\bea
\rho ~&=&~ \frac{N}{\pi \cosh x} ~+~ \delta \rho (x,t) ~, \nonumber \\
\theta ~&=&~ \frac{\hbar t}{8m} ~+~ \delta \theta (x,t) ~. 
\eea
The equations of motion then imply that
\bea
\frac{\partial \delta \rho}{\partial t} ~&=& ~- ~\frac{\hbar N}{\pi m} ~(~ 
\frac{1}{\cosh x} \frac{\partial^2}{\partial x^2} ~-~ \frac{\sinh x}{\cosh^2 
x} \frac{\partial}{\partial x} ~) ~\delta \theta ~, \nonumber \\
\frac{\partial \delta \theta}{\partial t} ~&=&~ \frac{\pi \hbar}{4Nm} 
\cosh x ~(~ \frac{\partial^2}{\partial x^2} ~+~ \frac{\sinh x}{\cosh x} 
\frac{\partial}{\partial x} ~+~ \frac{\sinh^2 x}{2 \cosh^2 x} ~+~ 
\frac{9}{2 \cosh^2 x} ~-~ \frac{1}{2} ~)~ \delta \rho ~.
\eea
To solve these equations, we eliminate $\delta \theta$ and define a function
$\psi (x)$ as
\beq
\delta \rho (x,t) ~=~ (\cosh x)^{-1/2} ~\psi (x) ~e^{-i \omega t} ~.
\label{psi}
\eeq
We then obtain the eigenvalue equation 
$$A ~\psi (x) ~=~ \omega^2 ~\psi (x), ~$$
where 
\beq
\quad A ~= ~\frac{\hbar^2}{4m^2} ~(\frac{d^2}{dx^2} ~-~ 
\frac{1}{4} ~+~ \frac{3}{4 \cosh^2 x})~ (\frac{d^2}{dx^2} ~-~ \frac{1}{4} ~
+~ \frac{15}{4 \cosh^2 x} ) ~\psi ~.
\label{eigen}
\eeq
The operator $A$ is hermitian with the weight $w (x) =1$; namely, for any two
normalizable functions $\psi_1 (x)$ and $\psi_2 (x)$,
\beq
\int_{-\infty}^\infty ~dx ~w(x) ~\psi_1^\star (x) ~A ~\psi_2 (x) ~=~
\int_{-\infty}^\infty ~dx ~w(x) ~(A ~\psi_1 (x))^\star ~\psi_2 (x) ~,
\eeq
with $w(x) =1$. (It is for this reason that we introduced the factor of
$(\cosh x)^{1/2}$ in the definition of $\psi$ in Eq. (\ref{psi})).
Hence the eigenvalues of $A$, $\omega^2$, must
be real. However, $\omega^2$ could be negative; in that case, there would
be a solution in which $\omega$ is purely imaginary and $-i\omega >0$. That
would be an unstable mode with the function $\delta \rho (x,t)$ blowing up 
exponentially as $t$ goes to $\infty$. We therefore need to know if 
$A$ has any negative eigenvalues.

We already know two eigenfunctions of $A$ with $\omega =0$ corresponding to 
the translation and scaling symmetries. From infinitesimal versions of these
two symmetries, we find that the eigenfunctions of the operator $A$ with
zero eigenvalues are given by 
\beq
\psi ~=~ \frac{\sinh x}{(\cosh x)^{3/2}} ~,
\eeq
and
\beq
\psi ~=~ \frac{1}{(\cosh x)^{1/2}} ~-~ \frac{x \sinh x}{(\cosh x)^{3/2}} ~.
\eeq
Note that these two functions are normalizable (i.e., they would be called 
bound states in the language of quantum mechanics), and that they have one 
and two nodes respectively.
Thus the operator $A$ has quite different properties from the usual 
Hamiltonian operator $H = -d^2/dx^2 + V(x)$ which appears in a 
one-dimensional Schr\"odinger equation; a second derivative operator like $H$
always has non-degenerate eigenvalues for bound states, and the corresponding
eigenvalues strictly increase with the number of nodes.

We have not been able to find an analytical expression for the eigenfunction
of $A$ corresponding to an eigenvalue $\omega^2 < 0$. However it is possible 
to show its existence by a variational argument. We use the result from 
quantum mechanics that for a hermitian operator $A$, if $\psi_v (x)$ is a 
normalizable variational function, and 
\beq
E_v ~=~ \frac{< \psi_v | A | \psi_v >}{< \psi_v | \psi_v >} 
\label{vare}
\eeq
is the corresponding variational energy, then there must be an eigenvalue of
$A$ which is less than or equal to $E_v$. We now try a function of the form
\beq
\psi_\alpha (x) ~=~ \frac{1}{(\cosh x)^{1/2}} ~+~ \alpha ~\frac{x \sinh 
x}{(\cosh x)^{3/2}} ~, 
\eeq
where $\alpha$ is a variational parameter which we assume to be real for
convenience. Then the variational energy given in (\ref{vare}) takes the form
\beq
E_\alpha ~=~ \frac{3\hbar^2}{10 m^2} ~\frac{\alpha (1+\alpha)}{1 + 2\alpha + 
\alpha^2 (1+ \pi^2 /8)} ~.
\eeq
We find numerically that this has a minimum value of $E=-0.1350 ~\hbar^2 /m^2$ 
at $\alpha = -0.4738$. Since this minimum value is negative, we know that 
there must be an eigenfunction of the operator $A$ for which $\omega^2 < 0$.
This shows that the bright soliton has at least one small fluctuation
mode which blows up as $t \rightarrow \infty$. Since we do not know the 
explicit form of this eigenfunction, the physical interpretation of this mode 
remains unclear. 

Although we have found an instability classically, it is possible that the
instability may disappear if the theory is quantized, i.e., if we impose the
equal-time commutation relation $[\rho (x), \theta (y)] = i \delta (x-y)$ 
\cite{sen2}. There are examples known in the literature (for instance, 
skyrmions in three dimensions) where the quantization of some collective 
coordinates eliminates a classical instability \cite{abdalla}. 
However, we will not examine the question of the quantum stability of the 
bright soliton in this paper where our entire analysis has been classical.

\section{General treatment of stationary solutions}

We will now see how the general form of the stationary solution of Eq. 
(\ref{scheq}) may be found. We do this by mapping the one-dimensional 
problem considered in previous sections to that of a pseudo-particle 
problem in two dimensions. To do this let us write 
\beq \Phi (x,t) = e^{-i\mu t /\hbar} ~[~ \xi_1 (x) ~+~ i \xi_2 (x) ~] ~.
\eeq
Thus the density given by $\rho = \xi_1^2 + \xi_2^2$ is independent of
time. Eq. (\ref{scheq}) then takes the form
\bea
\frac{d^2 \xi_1}{dx^2} ~&=&~ -~ \frac{2m\mu}{\hbar^2} ~\xi_1 ~ +~ g \pi^2 
\rho^2 ~\xi_1 ~, \nonumber \\
\frac{d^2 \xi_2}{dx^2} ~&=&~ -~ \frac{2m\mu}{\hbar^2} ~\xi_2 ~ +~ g \pi^2 
\rho^2 ~\xi_2 ~.
\label{eom1}
\eea
These are precisely the equations of motion of a classical particle of unit 
mass moving in two dimensions with the spatial coordinates $(\xi_1 , \xi_2 )$ 
and a {\it time} coordinate $x$. Note that in the one-dimensional problem 
$x$ refers to position coordinate. The total energy and pseudo-angular 
momentum of the particle are given by 
\bea
{\cal E} ~&=&~ \frac{1}{2} ~\left( \frac{d \xi_1}{dx} \right)^2 ~+~ 
\frac{1}{2} ~\left( \frac{d \xi_2}{dx} \right)^2 ~+~ \frac{m\mu}{\hbar^2} ~
\rho ~-~ \frac{g\pi^2}{6} ~\rho^3 ~, \nonumber \\
L ~&=&~ \xi_1 ~\frac{\partial \xi_2}{\partial x} ~-~ \xi_2 ~\frac{\partial 
\xi_1}{\partial x} ~.
\label{enang}
\eea
The pseudo-angular momentum $L$ is in fact the same, up to a factor, as 
the momentum density given in Eq. (\ref{mom}).
These are conserved since the particle is moving in a central potential
\beq
V (\xi_1 , \xi_2 ) ~=~ \frac{m\mu}{\hbar^2} ~\rho ~-~ \frac{g\pi^2}{6} ~
\rho^3 ~.
\label{pot}
\eeq
Since there are two conserved quantities and the particle has two degrees of 
freedom, the motion can be found exactly if the initial conditions are given.
For instance, $\rho$ is given in terms of $x$ by the equation
\beq
\left( \frac{d\rho}{dx} \right)^2 ~=~ 4 ~[~ 2 \rho {\cal E} ~-~ L^2 ~-~
\frac{2m \mu}{\hbar^2} ~\rho^2 ~+~ \frac{g\pi^2}{3} ~\rho^4 ~]~.
\label{eom2}
\eeq

We are interested in solutions of Eqs. (\ref{eom1}) and (\ref{eom2})
in which the particle density $\rho$ remains finite as $x \rightarrow 
\pm \infty$. In that case, the solution for $\rho$ will generally be a 
periodic function of $x$; this corresponds to a stationary wave. From 
the form of the potential energy in (\ref{enang}),
we see that this will happen in the following situations.

\noindent (i) If $g>0$, the potential has a negative sixth-degree term. 
Hence we must have $\mu > 0$ and the particle energy $\cal E$ must be
equal to or less than the potential barrier (this can be derived from Eq.
(\ref{pot})) in order to have $\rho$ finite for all times $x$. Then $\rho$
will always remain bounded by the location of the potential barrier given
by $\rho_0 = {\sqrt {2m\mu /g \pi^2 \hbar^2}}$ as in Eq. (\ref{tfd}). 

\noindent (ii) If $g<0$, the potential has a positive sixth-degree term, so 
$\rho$ will be finite regardless of whether $\mu$ is positive or negative. 
However the possible motions are different for $\mu >0$ and $\mu <0$ as we 
will see.

We will now discuss how the various solutions discussed in the previous 
section can be obtained as special cases from this general discussion. First 
of all, let us set the pseudo-angular momentum $L=0$ and therefore $\xi_2 =0$. 
Let us now consider the cases with $g$ positive and negative separately.

\noindent
(i) In the repulsive case $g>0$, we have already seen that we need $\mu >0$
and the energy must lie below the potential barrier so that the particle does 
not escape to infinitely large values of $\rho$. If the energy is 
exactly equal to the potential barrier, then we have two possible solutions. 
Either the particle stays at the top of the potential barrier for all times 
$x$ (this corresponds to the Thomas-Fermi result in (\ref{tfd})), or we have 
a solution in which the particle begins at $\xi_1 =-{\sqrt {\rho_0}}$ at time 
$x = - \infty$ and ends at $\xi_1 = {\sqrt {\rho_0}}$ at time $x = \infty$. 
The latter solution corresponds 
to the dark soliton of Kolomeisky {\it et al}. On the other hand, if the
energy is less than the potential barrier, then the particle will perform
a periodic motion which repeatedly passes through $\xi_1 =0$; this 
corresponds to a stationary wave with the minimum density being zero at an 
infinite number of values of $x$. This is a new stationary wave solution.

\noindent
(ii) In the attractive case $g<0$, there is only one kind of motion possible 
if $\mu >0$, i.e. the particle will perform a periodic motion passing 
repeatedly through $\xi_1 =0$, thus giving a new stationary wave. If $\mu <0$,
we have a double well potential (see Eq. (\ref{pot}), and there are various 
kinds of motion possible depending on the energy $\cal E$. If ${\cal E} > 0$, 
then the particle again performs a periodic motion passing through $\xi_1 =0$, 
thereby giving the stationary wave described in Sec. II C.
If ${\cal E} =0$, then we have a bounce solution in which the particle
begins at $\xi_1 = 0$ at $x = -\infty$, goes to some maximum (or minimum)
value of $\xi_1$ at some finite time and then returns to $\xi_1 =0$ at
$x = \infty$. This corresponds to the bright soliton described in Sec. II C;
note that such a solution necessarily has $N^2 g = -3/4$ no matter what the
value of $\mu$ is. (Thus there is no bright soliton solution possible if 
$N^2 g \ne -3/4$). If $\cal E$ is negative but is greater than the bottom 
of the potential, the particle performs a periodic motion in 
which $\xi_1$ always remains positive (or negative) and never reaches zero.
This is a new solution corresponding to a stationary wave whose minimum
density is nonzero. Finally, if $\cal E$ is equal to the bottom of the 
potential located at $\rho_0 = {\sqrt {2m\mu /g \pi^2 \hbar^2}}$, then we have
a solution with uniform density $\rho (x) = \rho_0$.

\section{Comments}

We have shown that the modified one-dimensional GP mean-field theory 
has many solutions apart from the one outlined by Kolomeisky {\it et 
al.} \cite{kolo}. The soliton solution in the attractive case is 
particularly simple since the full time dependent solutions in an 
arbitrary moving frame may be obtained simply by boosting the solutions 
in the stationary frame. The question of the quantum stability of this 
soliton remains an open problem.

S. G. thanks R. Rajaraman for discussions and comments. R.K.B.
acknowledges financial support from NSERC, Canada, for this research, and
the hospitality of I.M.Sc., Chennai, India.

\end{document}